\documentclass[10pt,preprintnumbers,amsmath,amssymb,prl,showpacs,twocolumn]{revtex4}
\usepackage{stmaryrd}
\usepackage{amsmath}
\usepackage{amssymb}
\usepackage{float}
\usepackage{array, color}
\usepackage{tabularx}
\usepackage{threeparttable}
\usepackage{titletoc}
\usepackage{booktabs}
\usepackage{graphicx}
\usepackage[colorlinks,linkcolor=blue,anchorcolor=blue,citecolor=blue,urlcolor=blue,dvipdfm]{hyperref}
\usepackage{graphics}
\usepackage{graphicx}
\usepackage{subfigure}

\usepackage{latexsym,bm}

\begin{document}

\title{Dynamical Ionic Clusters with Flowing Electron Bubbles from Warm to Hot Dense Iron along the Hugoniot Curve}

\author{Jiayu Dai$^1$} \author{Dongdong Kang$^1$} \author{Zengxiu Zhao$^1$} \author{Yanqun Wu$^2$} \author{Jianmin Yuan$^{1,}$}\email{jmyuan@nudt.edu.cn}

\address{$^1$Department of Physics, College of Science, National University of Defense Technology, Changsha 410073, P. R. China\\
$^2$College of Optoelectric Science and Engineering, National University of Defense Technology, Changsha 410073, P. R. China}

\date{\today}

\begin{abstract}
The complex structures of warm and hot dense matter are essential to understand the behaviors of materials in high energy density physics processes and provide new features of matter constitutions. Here, along a new unified first-principle determined Hugoniot curve of iron from normal condensed condition up to 1 Gbar, the novel structures characterized by the ionic clusters and separated "electron bubbles" are revolutionarily unraveled using newly developed quantum Langevin molecular dynamics (QLMD). Subsistence of complex clusters, with bonds formed by inner shell electrons of neighbor ions, can persist in the time length of 50 femto-seconds dynamically with quantum flowing bubbles, which are produced by the interplay of Fermi electron degeneracy, the ionic coupling and the dynamical nature. With the inclusion of those complicated features in QLMD, the present data could serve as a first-principle benchmark in a wide range of temperatures and densities.
\end{abstract}

 \maketitle
%\pacs{ 62.50.-p, 31.15.A-, 52.27.Gr, 64.30.-t} \maketitle
%05.70.Ce Thermodynamic functions and equations of state
%52.27.Gr Strongly-coupled plasmas
%71.15.-m Methods of electronic structure calculations
%64.60.Cn Order¨Cdisorder transformations
%52.65.Yy Molecular dynamics methods (in plasma)
%62.50.-p High-pressure effects in solids and liquids
%52.25.Kn Thermodynamics of plasmas
%62.50.-p High-pressure effects in solids and liquids
%31.15.A- Ab initio calculations (atoms and molecules)
%61.20.Ja, liquid structure
%64.30.-t Equations of state of specific substances
%64.30.Ef Equations of state of pure metals or alloy

The thermodynamic and structural properties of matters at extreme conditions, so-called warm dense matter (WDM) and hot dense matter (HDM) in the field of high energy density physics (HEDP) \cite{s2}, are both experimental and theoretical challenges and critical to the comprehension of the evolution and the internal structures of giant planets, stars, inertial confinement fusion (ICF) target capsule, and material science \cite{s2,s3,s4,s5,s6,s7,s8}. New physics discovered by high-power laser facilities such as National Ignition Facility (NIF) and nuclear fusion \cite{s10,s11,s12,s13,s14,s15} requires understanding beyond the traditional condensed matter and atom (or plasma) physics \cite{s2,s7}. Recent laser-driven dynamical experiments and related theories show the existence of ordered electron-ion structures \cite{s15,s16,s17,s18,s19,s20,s21,s22} and electronic bonds \cite{s13,s23} in WDM and HDM with x-ray methods, suggesting the need to take into account of the dynamics of local chemical environments. Meanwhile, studies of static high-pressure theories and experiments have led to the finding of "electron blobs" formed by valence electrons and new-type electronic bonds assisted by inner-shell electrons in cold aluminum and sodium at high pressures \cite{s24,s25}. However, for the lack of effective methods, few theoretical studies are carried out on the structures of complex materials in HEDP field, which are crucial for determining their physical properties such as energies, pressures and transport behaviors.

In the HDM, the densities are comparable to or even much higher than the states of static compression and WDM, and temperatures are up to hundreds of eV \cite{s2}. The higher density effect could induce new features that cannot be seen in the normal WDM, while the higher temperature causes dynamical changes of the ionic configurations accompanied with dynamical electronic distributions, which do not exist in the static highly compressed cold matters. The properties of these kinds of matters, including the equation of states (EOS), electronic and ionic conductivities, viscosities and diffusions, and optical properties, are definitely dependent on the unknown details of the electron-ion structures under these extreme conditions.

In order to shed light on the hidden features and controversial intrinsic dynamics from WDM to HDM, the electron-ion structures are calculated along the originally determined principal Hugoniot curve of Fe using newly developed first principles method-quantum Langevin molecular dynamics (QLMD) \cite{s26,s27}. For Fe, as one of the most abundant elements in universe and a typical complicated transitional metal, it is a long-standing challenge \cite{s4,s5,s6,s15,s28,s29,s30,s31,s32} to obtain the physical properties such as EOS and electron-ion structures, because of the strong ionic coupling, and high electronic degeneracy in a wide range of temperature and density. To date, previous experiments and statistical framework based theories generate abundant results with large divergence and uncertainty in EOS \cite{s28,s29,s30,s31,s32}. First principles studies on the EOS and electronic properties of crystalline or liquid Fe at high pressure and zero or relatively low temperature have been reported  widely \cite{s4,s5,s6,s33,s34}, and a few fixed density-temperature points on the Hugoniot curve picked up from SESAME table were calculated by quantum molecular dynamics (QMD) \cite{s27,s28}. However, none Hugoniot data beyond WDM from first principles are covered.

Hugoniot curves are determined by both the pressures and the internal energies. QLMD or QMD, based on finite-temperature density functional theory (DFT) \cite{s35}, can naturally include the effects of degeneracy and coupling contributing to the pressure and energy, and has been successfully applied to derive the EOS and dynamical properties of dense matter including Fe \cite{s3,s33,s34,s36,s37,s38,pimc}. Advantageously, QLMD, adopted in the Quantum Espresso package \cite{s47}, can be extended to the HEDP field within the framework of \textit{ab initio} by introducing electron-ion collision induced friction (EI-CIF) \cite{s27}. It is thus possible to accurately explore the details of the electron-ion structures in WDM and HDM as a powerful tool. In the present work, 54 atoms are included in the supercell with $3\times 3\times 3$ k-points below 10 eV and $\Gamma$ point only at higher temperatures for the representation of Brillouin zone. Pseudopential (PP) with 16 electrons in the valence within the generalized-gradient approximation (GGA) \cite{s48} is used, since the ionization degrees are less than 16 below 100 eV \cite{s27}. During the MD processes, the time steps are from 1 $fs$ to 0.25 $fs$ with increasing the temperature, and 2 $ps$ time length is chosen to achieve the thermal stability state. After the thermalization, more than 2 $ps$ time length is used to acquire the thermal properties such as EOS and ionic structures. Over 300 density-temperature points are calculated in order to get the EOS data.

\begin{figure}[!tb]
\centering
\includegraphics*[width=3.2in]{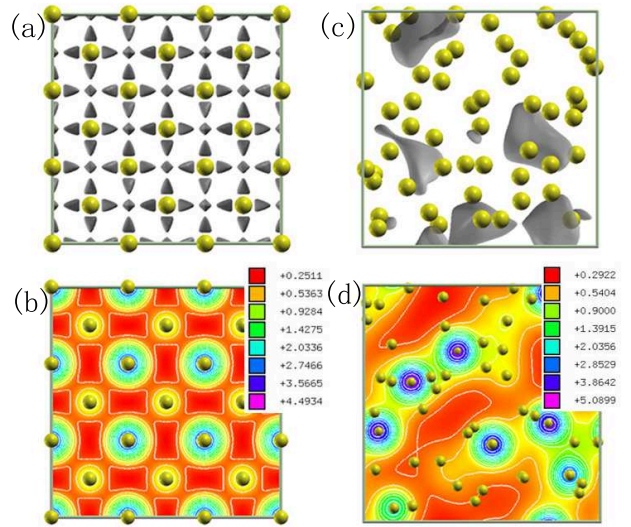}
\caption{(Color) The electronic charge density (electron/{\AA}$^3$) distributions of Iron. (a) and (b): the three and two dimensional in the (100) direction contour plot for the charge density of Iron in bcc phase at (0 eV, 33.385 g/cm$^3$); (c) and (d): the same contour plot of Iron at (100 eV, 33.385 g/cm$^3$).} \label{Fig1}
\end{figure}

The EOSs of the temperatures from 0.1 eV to 100 eV and pressures up to 1 Gbar (1 Gbar = 100 TPa) on the both sides of the Hugoniot curve are obtained. We firstly discuss the electron-ion structures and the dynamics along this curve, and finally give the EOS data from normal condensed state to WDM and to HDM. One of the most spectacular physics here is the electronic structures in WDM and HDM, which we knew few today. To dig out the undiscovered features, the electronic distributions at the density-temperature point of (100 eV, 33.385 g/cm$^3$) on the calculated Hugoniot curve are displayed. It is shown in Fig.~\ref{Fig1}(a) that the formation of "blobs" of the valence electrons of cold iron at high pressure, which was proved in high density cold aluminum \cite{s24}. The contraction at high pressure makes the ions close enough, with the free-like valence electrons occupying their left space. When the valence electrons become free, the inner \textit{s}, \textit{p}-electrons will assist the bonding on Fe-Fe, as shown in the band structures in very recent high-density results \cite{s33}. This can be verified in the 2-dimensional density distribution in Fig.~\ref{Fig1}(b), where the valence electrons (low densities) are distributed between Fe ions, and the inner electrons form covalent bonds. How can this feature change when the dynamical effects are introduced? As shown in Fig.~\ref{Fig1}(c), the valence electron "blobs" tend to assemble together and form bigger "bubbles" in the interspaces of Fe ions due to the temperature induced dynamics. These free electrons are not homogenous and behave as "quantum electron liquids" flowing with ionic moving, which can not be described by the current atom and plasma models. It can be verified in the 2-dimensional picture in Fig.~\ref{Fig1}(d), where the free electrons (red color) are distributed in the interspaces of ions. Interestingly, there are clear-cut density overlaps among more than three ions at high density induced by the inner orbital electrons, indicating the existence of many-body bonding formed by inner shell electrons. Furthermore, the covalent-like bonds are shown, which are likely formed by the \textit{s} to \textit{p} or \textit{sp} to \textit{d} electron transfer and their hybridization at high density \cite{s33}.

The "electron bubbles" due to the interplay of the Fermi electron degeneracy, the ionic coupling and temperature-induced dynamics are obviously separated from the ions, which can be considered as two phases coexisting and flowing in HDM. This discovery is quite different from the physics in WDM and ideal plasma gases. In WDM, the electrons and ions are similar to the physics in condensed matter, where the electrons "move" clingingly to ions "simultaneously" and distribute around some specific atoms. In ideal plasma gases, the electrons and ions are all considered as classical non-interacting or weakly coupled particles. Here in dense and hot matter, the ions can form networks distorting slowly in the quantum electron fluid, giving an alternative quantum mechanical case for the two phases. This feature with bonds should not be found in the simple elements such as hydrogen, whose electronic structures are much simpler, and there are no "inner-shell electrons" assisting the formation of new bonds and excluding the ionized valence electrons from the region of gathered nuclei due to the Fermi degeneracy.

\begin{figure}[!tb]
\centering
\includegraphics*[width=3.1in]{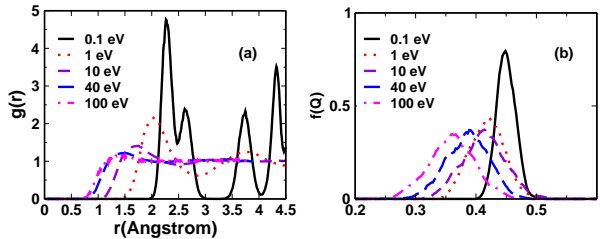}
\caption{(Color online) (a) RDF and (b) the distribution of orientation order parameters Q for the selected five temperature-density points along Hugoniot curve. The corresponding temperatures and densities are respectively (0.1, 1, 10, 40, and 100) eV, and (10.1, 13.23, 18.75, 28.875, and 33.385) g/cm$^3$.} \label{Fig2}
\end{figure}

The electronic distributions show the complexity of the ionic structures, whose details and dynamics are still secrets. Most importantly, the electronic structures are tightly dependent on the dynamics of ions and their collective behaviors. In order to understand the physics of structures, we select five temperature-density (T-D) points as shown in Fig.~\ref{Fig2} along the new principal Hugoniot curve to uncover their hidden  dynamic nature of the ionic structure. Their radial distribution functions (RDF) shown in Fig.~\ref{Fig2}(a) give evidences of the transition of ionic structures from long-range order to short-range one statistically. It is worth noting that even at T = 100 eV, there is also one peak in RDF, indicating the existence of hidden ordered structures.

Considering the short-range ordered structures at high temperatures, we borrow the language of the liquid structures such as water and clusters to reveal the structures in HEDP. The orientation order parameter $Q=1-\frac{3}{8}\sum_{i=1}^3\sum_{j=i+1}^4(\cos\theta_{ij}+\frac{1}{3})^2$ is defined, where $\theta_{ij}$ is the angle formed by the lines of an ion and its nearest neighbors $i$ and $j$ ($\leq4$). The value of Q varies from 0 (in an ideal gas) to 1 (in a perfect tetrahedral network), which can be used as a measure of tetrahedrality for the local coordination structure \cite{s39,s40}. As shown in Fig.~\ref{Fig2}(b), the peak of the distribution of the parameter Q shifts from 0.45 to 0.35 with increasing temperature, indicating that the ionic structures with the ideal tetrahedral network collapse indeed but there are still some similar topological structures at high temperatures.

As for the topological structures, some ordered structures survive even at 100 eV from the hints of Fig.~\ref{Fig1} and Fig.~\ref{Fig2}. However, there is not a clear minimum after the first peak in the RDF in Fig.~\ref{Fig2}(a) except for 0.1 eV. Therefore, a dissociation criterion based solely on a hard cutoff on Fe-Fe bond lengths would be optional. The probability distribution of coordination numbers (CNs) can be good for analyzing the local geometry, which has been successfully used for estimating the dissociation for water molecules at high temperature \cite{s41}. Due to the short ordered structures in HDM \cite{s42}, here we adopted the effective CNs (ECNs) idea\cite{s43,supp}. For low-symmetry structures where a particular atom is surrounded by atoms at different distances, the ECNs concept can be independent on the choice of the bond cutoff, and therefore provides a more accurate method to determine possible structural trends in disordered structures. Thus, ECNs can lead to more clearly structural figures.

For the above five T-D points, ECNs decrease from 10.99 to 2.93 {\AA}, whose corresponding average bond lengths are from 2.33 to 1.15 {\AA}. With increasing temperatures and densities, ECNs decrease gradually, but retain larger than the unit even up to 100 eV. This fact indicates the existence of cluster-like or network-like structures, which is consistent with the hints in Fig.~\ref{Fig1}. A part of ions catch only one nearest neighbor (ECN = 1) when temperature is high enough, indicating the formation of two-ion chains. The distributed percentages of these "chains" are 13\%, 26\%, and 36\% respectively for the temperatures of 10, 40 and 100 eV. Furthermore, the differences between the dynamical average bond lengths and ionic radii (from 2.10 to 1.41 {\AA}) show that the interatomic distances can not be simply described by the hard sphere model, and the average sphere space can not be only responsible for the density.

\begin{figure}[!tb]
\centering
\includegraphics*[width=3.1in]{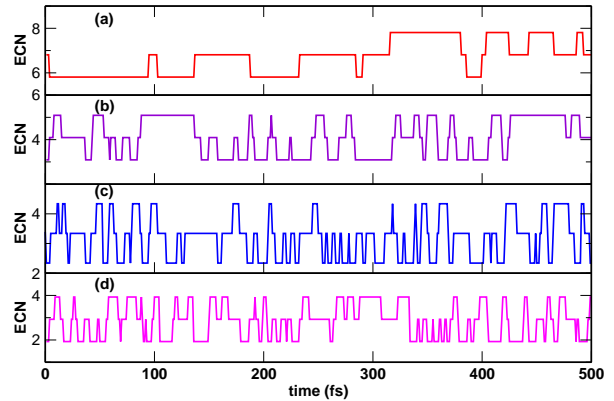}
\caption{(Color online) The trajectories of ECNs for a specific atom during the time length of 0.5 ps simulations at different states of (a), 0.1 eV, (b), 1 eV, (c), 10 eV, and (d), 100 eV, respectively.}
\label{Fig3}
\end{figure}

The dynamics of these bond-network breaking and forming patterns, i.e., the dynamical dissociation of the Fe-Fe bonds, is also pivotal since an instantaneous topological structure can not affect the observed physical properties obviously. In order to investigate the structural dynamics, we firstly trace the topological networks around one specific atom, revealing the dynamic nature of the topological structures during the simulations, shown in Fig.~\ref{Fig3}. With the increasing temperatures, the movements of ionic positions in liquid states introduce some physical processes such as dissociations \cite{s41}, resulting in the change of structures at different times. However, even though the structures exhibit dynamical changes, a considerable fraction of ions in warm and hot dense Fe with compact clusters (networks) can persist for a long time of a few tens of femto-seconds (\textit{fs}) even at T = 100 eV. Because most values of ECN$_i$ (ECN$_i$ represents the ECN of the $i_{th}$ atom) are located around ECN within the departure of 2, only three patterns of topological structures are statistically averaged: using ECN as the reference, for the $i_{th}$ atom, we assume ECN$_i$=ECN-1 when ECN$_i$$\leq$ECN-1, and ECN$_i$=ECN+1 when ECN$_i$$\geq$ ECN+1, and ECN$_i$=ECN when ECN-1$<$ECN$_i<$ECN+1. Summing up the persisted time length for every structure, we can find that more than 15\% of the topological structures forming and breaking up on the time scale of longer than 20 \textit{fs} at 100 eV (Figure S3 in the supplementary \cite{supp}). With the moving cluster-like ions, "electron bubbles" will transport between different clusters' interspaces as flowing fluid. These bubbles may crash at one time, but will appear in the next time with different configurations in different interspaces. The dynamic natures will change rapidly, but the clusters will persist long enough at the time scale and high enough at the percentage to affect the electronic structures and related properties such as total energies, optical properties, and electronic conductivities \cite{s42}. This understanding violates the traditional assumption that the states at so high temperature can be modeled on the single atomic scale, and the collective and quantum essentials must be treated accurately from first principles. Furthermore, the dynamical behaviors of the topological structures introduce a challenge for the statistical models such as hypernetted-chain (HNC) \cite{s18,s19} to include more topological networks in its constructed potentials. From another point of view, the dominant two-ion chain structures at the temperature of 100 eV would induce closer pressures between the statistical methods and QMD (QLMD) \cite{s27,s28} methods due to the very simple structures.

\begin{figure}[!tb]
\centering
\includegraphics*[width=3.1in]{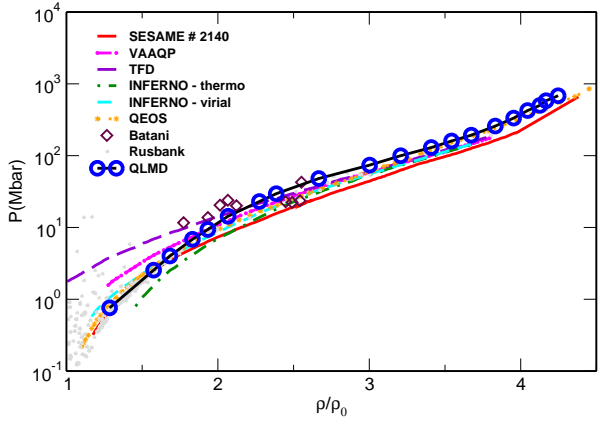}
\caption{(Color online) Principal Hugoniot of Fe, comparing the results from QLMD with other models and experiments. The corresponding temperatures are T = (0.1, 0.5, 1.0, 2.0, 3.0, 5.0, 8.0, 10.0, 15.0, 20.0, 25.0, 30.0, 35.0, 40.0, 50.0, 60.0, 70.0, 80.0, 90.0, and 100.0) eV.} \label{Fig4}
\end{figure}

Let us back to the principal Hugoniot curve in Fig.~\ref{Fig4}, which is determined by interpolating a few density points at a fixed temperature according to the Hugoniot-Rankine (HR) relation \cite{s36,s44}: $(U-U_0)=\frac{1}{2}(P+P_0)(V_0-V)$, where $U$, $P$ and $V$ are the internal energy, the total pressure and the volume of the system, respectively; $U_0$, $P_0$, $V_0$ are the respective parameters of the initial reference state (7.86 g/cm$^3$ and 20 K). 20 Hugoniot temperature-density points are determined up to the pressure of 1 Gbar, as shown in Fig.~\ref{Fig4}. It is noticeable that the spin-polarization is important at T = 0.1 eV.

The previous data of different experiments are scattered, which is mainly caused by the uncertainties of the temperature and density \cite{s6,s30}. Furthermore, different models give different results from WDM to HDM, especially in the WDM under the pressure of a few Mbar. In this regime, our first-principle results are along the lower limit envelop of the distributions of the experimental data, similar to SESAME table. With the increasing pressures above 10 Mbar, the Hugoniot curve derived from QLMD is very close to the experiments from rusbank data \cite{s29} and within the error bars of Batani's experiment \cite{s30}. In experiments, external factors such as preheating can affect the final results significantly \cite{s30}, inducing unexpected higher pressures. The SESAME tables are far from both experimental and our data at relatively high temperature, also shown by previous studies \cite{s30,s45}. According to the present calculations, it can be concluded that below 10 Mbar, the experimental results along the lower envelop would be regarded error-free, and the limit of pressures in experiments above 10 Mbar should locate around our points. Similar conclusions can be found in the Hugoniot of Al \cite{s36}, verifying the accuracy of our calculations. The statistical methods such as Thomas-Fermi-Dirac (TFD) \cite{s28}, variational-average-atom-in-quantum-plasmas (VAAQP), INFERNO \cite{s31} and quotidian equation of state (QEOS) \cite{s46} models do not give the same accurate results as QLMD,  mainly because physical quantities such as the energies are sensitive to many-body interactions and collective quantum electronic distributions and their dynamics at high densities and high temperatures, which are not fully considered in the above statistical models. The internal energies from QLMD should be lower than those of statistical models according to the variational principle. According to HR relation, lower internal energies at a definite pressure would result in lower densities. Therefore, the Hugoniot curve derived from QLMD would be higher than those of the others, which is the reason for the results of harder compression from first principles \cite{s3,s38}.

Finally, we construct the formula of all calculated EOS, i.e., in the range of $0.5\leq T \leq 100$ eV, $9\leq \rho \leq 45$ g/cm$^3$. This study is important for the applications of our data to the experimental benchmark and astrophysical hydrodynamics within the wide range of pressures. The optimal fit formula in a least-square sense is as follows according to the Virial expansions using a typical EOS relation:
\begin{equation}\label{eos}
P(T,\rho)=\sum_{m=0}^{M}(\sum_{n=0}^{N}\alpha_{mn}(log_{10}T)^n)\rho^m,(M=2, N=12)
\end{equation}
Where the units of P, T and $\rho$ are respectively kbar, K and g/cm$^3$. The values of the coefficients $\alpha_{mn}$ and the validation are shown in the supplementary materials  \cite{supp}.

In conclusion, the free electron "blobs" in dense matter move and assemble together, forming bigger "bubbles" separated from bond-network clusters at high temperatures, behaving like quantum flow. The dynamical ionic structures are analyzed according to the topological structures based on the ECNs idea, giving a new realization of the stable existence of compact clusters contributed by the inner-shell electrons forming bonds even at T = 100 eV. The ionic structures with bound electrons can be compared to the soft "skeleton" in the system, which would be responsible for the shearing strength, viscosity and response to the shock waves; while the free electrons can be compared to "protoplasm" flowing among the skeletons contributing to kinetic pressure of electrons, and conductivities. In addition, these unique features would raise a challenge to understand the dynamical formation of ordered structures from non-equilibrium to equilibrium. At last, the new Hugoniot data from QLMD simulations give a benchmark on the EOS of Fe, which can be regarded as the converged limit of the error-free experimental data.

This work is supported by the National NSFC under Grant Nos. 60921062 and 11104351. Calculations are carried out at the Research Center of Supercomputing Application, NUDT.

\end{document}